\documentclass[twocolumn,preprintnumbers,amssymb,amsmath,aps,floatfix,prc,superscriptaddress]{revtex4}

\usepackage{epsfig}
\usepackage{bm}
\usepackage{amssymb}
\usepackage{amsmath}
\usepackage{color}
\usepackage{subfigure}
\usepackage{hyperref}

\begin{document}
	
\title{Probing medium response via strangeness enhancement around quenched jets}

	\author{Ao Luo}
	\affiliation{Institute of Frontier and Interdisciplinary Science, Shandong University, Qingdao, Shandong 266237, China}
	
	\author{Shanshan Cao}
	\email{shanshan.cao@sdu.edu.cn}
	\affiliation{Institute of Frontier and Interdisciplinary Science, Shandong University, Qingdao, Shandong 266237, China}
	
	\author{Guang-You Qin}
	\email{guangyou.qin@mail.ccnu.edu.cn}
	\affiliation{Institute of Particle Physics and Key Laboratory of Quark and Lepton Physics (MOE), Central China Normal University, Wuhan, 430079, China}
	
\begin{abstract}

Jet-induced medium excitation is a crucial part of jet interactions with the quark-gluon plasma (QGP) in relativistic heavy-ion collisions, and has recently been confirmed by experiment for the first time.
Based on the AMPT model simulation, we propose the strangeness enhancement around quenched jets as a novel signature of jet-induced medium excitation.
By applying the jet-particle correlation techniques, we calculate jet-induced particle yields around the jets and find a significant enhancement of the strange-to-non-strange-hadron ratio and the double-to-single-strange-hadron ratio correlated with jets in relativistic nucleus-nucleus collisions relative to proton-proton collisions.
This enhancement increases with both the strength of jet-QGP interactions and the radial distance from jet axis.
These observations align with the features of jet-induced medium excitation and parton coalescence in hadron formation, and await experimental validation in the future measurements.
		
	\end{abstract}
	\maketitle

\section{Introduction}
\label{sec:introduction}
	
Energetic nuclear collisions conducted at the Relativistic Heavy-Ion Collider (RHIC) and the Large Hadron Collider (LHC) provide a unique opportunity for creating an extremely hot environment, similar to the early universe microseconds after the Big Bang, in which nuclear matter exists in a color deconfined state known as the quark-gluon plasma (QGP)~\cite{Gyulassy:2004zy,Jacobs:2004qv,Busza:2018rrf,Elfner:2022iae}. Jet is a spray of collimated particles developed from a highly virtual parton produced by the initial nucleon-nucleon scatterings. Due to interactions between jet partons and the QGP, the transverse momentum ($p_\mathrm{T}$) spectra of jets observed in nucleus-nucleus (A+A) collisions can be significantly suppressed relative to those in proton-proton ($p+p$) collisions. This phenomenon, called jet quenching, serves as a powerful probe of the QGP properties~\cite{Wang:1992qdg,Bass:2008rv,Qin:2015srf,Majumder:2010qh,Blaizot:2015lma,Cao:2020wlm}. Over the past three decades, significant efforts have been devoted to understanding the dynamics of jet quenching~\cite{Qin:2010mn,Majumder:2011uk,Dai:2012am,Blaizot:2013hx,Chien:2016led,Caucal:2018dla,Xing:2019xae,Huss:2020dwe,Mehtar-Tani:2021fud,Zhao:2021vmu} and developing sophisticated Monte-Carlo event generators to simulate jet-QGP interactions~\cite{Cao:2024pxc,Schenke:2009gb,Zapp:2013vla,Casalderrey-Solana:2014bpa,Cao:2017qpx,JETSCAPE:2017eso,Putschke:2019yrg,Luo:2023nsi,Karpenko:2024fgg}. Systematic comparisons between these model calculations and experimental data on jet quenching have offered valuable insights into the QGP properties, such as the jet transport coefficient inside the QGP~\cite{JET:2013cls,JETSCAPE:2021ehl,Xie:2022ght,Chen:2024epd}, its geometry~\cite{Karmakar:2024jak}, its specific viscosity~\cite{Karmakar:2023ity}, and its equation of state~\cite{Liu:2023rfi}.
	
Jet-medium interactions include not only medium modification on jets, such as jet energy loss and momentum broadening, but also jet modification on the medium by depositing (depleting) energy into (out of) the medium, which is referred to as jet-induced medium excitation, or medium response~\cite{Cao:2022odi}. Since there exists no perfect method to determine whether a (soft) hadron is emitted from the QGP background or produced from hard parton fragmentation in heavy-ion experiments, medium response is inevitably embedded in all jet observables, such as its nuclear modification factor~\cite{He:2018xjv,JETSCAPE:2022jer}, collective flow coefficients~\cite{He:2022evt}, and intra-structures~\cite{Casalderrey-Solana:2016jvj,Tachibana:2017syd,KunnawalkamElayavalli:2017hxo,Milhano:2017nzm,Chen:2020tbl,Park:2018acg,Luo:2018pto,Casalderrey-Solana:2019ubu,Chang:2019sae,Tachibana:2020mtb,Yang:2023dwc,Xing:2024yrb}. Searching for unique signatures of medium response is among central objectives of jet studies in heavy-ion programs. Since jets travel with a supersonic velocity, a Mach-cone structure of the wave front of medium response has been proposed in Refs.~\cite{Casalderrey-Solana:2004fdk} and later extensively investigated in Refs.~\cite{Chaudhuri:2005vc,Ruppert:2005uz,Gubser:2007ga,Chesler:2007an,Qin:2009uh,Neufeld:2009ep,Li:2010ts,Ma:2010dv,Casalderrey-Solana:2020rsj}. The detailed patterns of the jet-induced Mach-cone is expected to be sensitive to the speed of sound and the specific viscosity of the QGP~\cite{Neufeld:2008dx,Bouras:2014rea}.
Unfortunately, the Mach-cone structure of the wave front can be easily distorted by the radial flow of the QGP~\cite{Ma:2010dv,Betz:2010qh,Tachibana:2015qxa}, and may also be spoiled by interference between energy deposition from multiple shower partons within a jet~\cite{Neufeld:2011yh} and event-by-event fluctuations~\cite{Renk:2013pua}, and therefore is difficult for a direct observation. On the other hand, the energy depletion in the diffusion wake left behind jets, as proposed in Refs.~\cite{Betz:2008ka,Tachibana:2017syd,Chen:2017zte,Tachibana:2020mtb,Casalderrey-Solana:2020rsj,Yang:2021qtl,Yang:2022nei} and recently confirmed by experiment~\cite{CMS:CMS-PAS-HIN-23-006}, serves as an unambiguous footprint of medium response inside the QGP. This energy depletion, propagating in the direction opposite to a jet, arises from the energy conservation of the medium as part of the medium is driven by the jet along its trajectory, and thus can be hardly explained by jet theories without considering medium response.

Apart from the energy depletion in the diffusion wake, another promising signal of medium response could be the medium-modified hadron chemistry around jets.
Since the energy lost from jet can be further thermalized via interacting with medium constituents, the particles produced from jet-deposited energy are expected to be richer in baryons than those in $p+p$ collisions.
The enhancement of baryon-to-meson ratio around jets in A+A {\it vs.} $p+p$ collisions has been proposed by several model calculations~\cite{Chen:2021rrp,Luo:2021voy,Sirimanna:2022zje}, and a hint has also been seen in experiment~\cite{Dale-Gau:2023ree}. Similar to baryon enhancement, one can also expect an enhancement of strangeness production around the quenched jets due to medium response, considering the richer strangeness from thermal production inside the QGP than that produced in $p+p$ collisions~\cite{Rafelski:1982pu}. The abundance of strange hadrons, especially multi-strange hadrons, formed via coalescence between strange and non-strange quarks (or multiple strange quarks)~\cite{Fries:2003kq,Fries:2003vb,Greco:2003mm}, has been considered a crucial signature of the formation of the color deconfined QGP in energetic heavy-ion collisions. The goal of our present work is to explore the variation of strangeness abundance around jets resulting from jet-induced medium excitation together with coalescence between jet shower partons and medium partons.

We will use the AMPT model~\cite{Lin:2004en,Zhang:2005ni} to simulate the production of jets and the QGP medium, interactions between jet shower partons and the medium constituent partons, and the hadronization of partons. The distribution of jet-correlated hadrons with respect to the jet direction will then be calculated for both Pb+Pb and $p+p$ collisions at $\sqrt{s_\mathrm{NN}} = 5.02$~TeV. We will show that the strange-to-non-strange-hadron-ratio and double-to-single-strange-hadron-ratio of jet-correlated hadrons can be significantly enhanced in Pb+Pb collision relative to $p+p$ collisions.
This enhancement is more prominent in more central collisions and at larger distance from the jet axis, both in line with the features of medium response.

\section{Jet-hadron correlations}
\label{sec:correlation}

We utilize the AMPT model~\cite{Lin:2004en,Zhang:2005ni,Ma:2010dv} to simulate the formation, dynamical evolution and hadronization of both the QGP medium and high energy particles (jets).
It has been successfully applied to studying a variety of phenomenologies in heavy-ion collisions, such as the anisotropic collective flow coefficients of soft hadrons~\cite{Lin:2001zk, Chen:2004dv, Xu:2011fe, Ma:2010dv}, dijet and $\gamma$-jet asymmetries~\cite{Ma:2013pha, Ma:2013bia}, jet fragmentation function~\cite{Ma:2013gga}, jet shape~\cite{Ma:2013uqa}, and enhancement of baryons around quenched jets~\cite{Luo:2021voy}. Following Ref.~\cite{Luo:2021voy}, we explore the jet-induced strangeness production as jets traverse the QGP.
		
In the string melting version of the AMPT model, hadrons produced from the HIJING model~\cite{Wang:1991hta, Gyulassy:1994ew} melt into their constituent quarks and antiquarks, providing the initial state of the subsequent partonic cascade. The initial geometry of nuclei, hard and soft scatterings between nuclei, and the cold nuclear matter effect on jet production are all included in the HIJING model. To ensure efficiency in generating high $p_\mathrm{T}$ jets,  the jet trigger technique in HIJING is implemented in our AMPT simulation. Quarks (and antiquarks) created by string melting then scatter with each other via the ZPC module~\cite{Zhang:1997ej}, which tracks the evolution of partons in the phase space based on two-body elastic scatterings. The partonic cross-section of these scatterings are set as $1.5$~mb in our following calculations for Pb+Pb collisions at $\sqrt{s_\mathrm{NN}} = 5.02$~TeV, which provides reasonable descriptions of the nuclear modification factor of jets and dijet asymmetry in earlier AMPT simulations~\cite{Gao:2016ldo, Luo:2021hoo}. At the end of their partonic scatterings, quarks are recombined into hadrons based on the quark coalescence mechanism~\cite{Lin:2001zk}. In AMPT, two (three) quarks close to each other in the coordinate space are selected to form a meson (baryon). Once hadrons are formed, they undergo further interactions via the ART module~\cite{Li:1995pra}, which includes both elastic and inelastic scattering processes until kinetic freeze-out of the system or the maximum scattering time for hadrons (set as 30~fm) is reached.
	
To extract the distribution of identified particles around jets from the AMPT data, we employ the jet-particle correlation method that has been widely applied on the experimental data~\cite{CMS:2016cvr, CMS:2016qnj, CMS:2018zze, CMS:2021nhn}.
We first construct the two-dimensional density distribution of identified particles ${d^2N}/({d\Delta\phi d\Delta\eta})$, where ${\Delta \eta}$ and ${\Delta \phi}$ are the relative pseudo-rapidity and azimuthal angle between associated particles and triggered jet. In this work, jets are reconstructed using the Fastjet package~\cite{Cacciari:2008gp, Cacciari:2011ma}, in which the anti-$k_\mathrm{T}$ algorithm is selected and a cone size of $R=0.4$ is set. Particles that satisfy $|\eta| < 2.4$ and $0.4 < p_\mathrm{T} < 300$~GeV are used for jet reconstruction.
The medium background is subtracted using the area based method, where the background density of an event is estimated using the median value of densities of $k_\mathrm{T}$-jets in the event, with two hardest jets excluded.
After background subtraction, jets with $p^{\rm jet}_\mathrm{T} > 120$~GeV and $| \eta_{\rm jet} | < 1.6$ are selected for further analysis, and identified particles are required to satisfy $|\eta| < 2.4$. These pseudorapidity cuts on jets and identified particles lead to a non-uniform acceptance rate with respect to $\Delta \eta$. To account for this limited acceptance effect, a mixed-event method is applied to correct the jet-particle correlations as~\cite{CMS:2016cvr, CMS:2016qnj, CMS:2018zze, CMS:2021nhn}
	\begin{align}
		\frac{1}{N_{\rm jet}} \frac{d^2N}{d{\Delta\eta} d{\Delta\phi}}
		= \frac{ME(0,0)}{ME({\Delta \eta},{\Delta\phi})} S({\Delta \eta},{\Delta\phi}).
		\label{eq:dN_per_jet}
	\end{align}
In the equation above, $S({\Delta \eta},{\Delta \phi})$ denotes the signal jet-particle pair distribution from the same event, while $ME({\Delta \eta},{\Delta \phi})$ denotes the mixed-event pair distribution in which identified particles are taken from a different event than the one containing a given jet. After being normalized to per trigger jet, they are written as
	\begin{align}
		S({\Delta \eta},{\Delta\phi}) &= \frac{1}{N_{\rm jet}} \frac{d^2N^{\rm same}}{d{\Delta\eta} d{\Delta\phi}},
		\nonumber\\
		ME({\Delta \eta},{\Delta\phi}) & =
		\frac{1}{N_{\rm jet}} \frac{d^2N^{\rm mixed}}{d{\Delta\eta} d{\Delta\phi}}.
	\end{align}
The factor ${ME(0,0)}/{ME({\Delta \eta},{\Delta\phi})}$ in Eq.~(\ref{eq:dN_per_jet}) then corrects the limited acceptance effect on $S({\Delta \eta},{\Delta\phi})$.

\begin{figure}
		\includegraphics[width=0.99\linewidth]{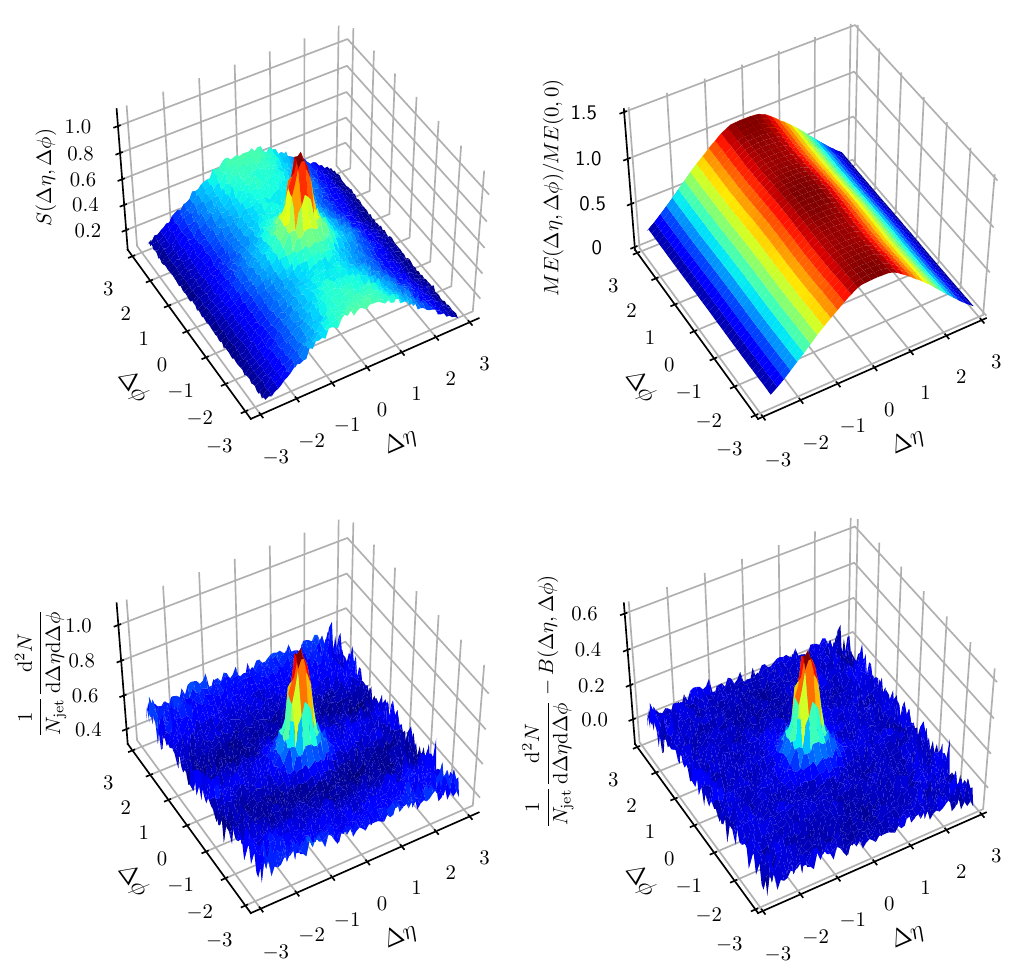}
		\caption{(Color online) Jet-kaon correlations for associated kaons within $2<p_\mathrm{T}<3$~GeV in 0-30\% Pb+Pb collisions at $\sqrt{s_\mathrm{NN}} = 5.02$~TeV: upper left for the signal pair distribution, upper right for the central-value-normalized mixed-event pair distribution, lower left for the acceptance-corrected distribution, and the lower right for the final distribution of jet-correlated kaon yield after background subtraction.}
		\label{fig:mixed_event}
\end{figure}
	
\begin{figure*}[t!]
		\includegraphics[width=0.49\linewidth]{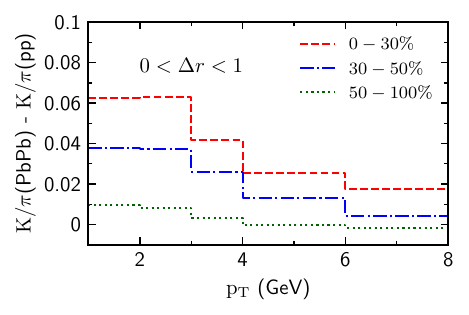}
		\includegraphics[width=0.49\linewidth]{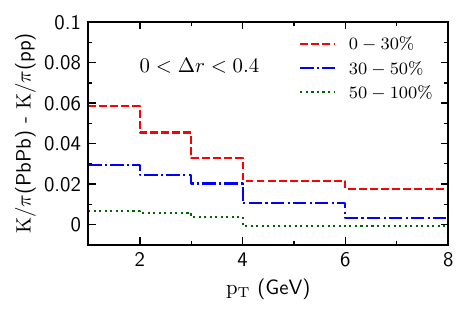}
		\includegraphics[width=0.49\linewidth]{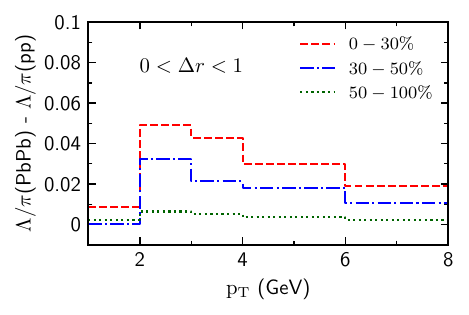}
		\includegraphics[width=0.49\linewidth]{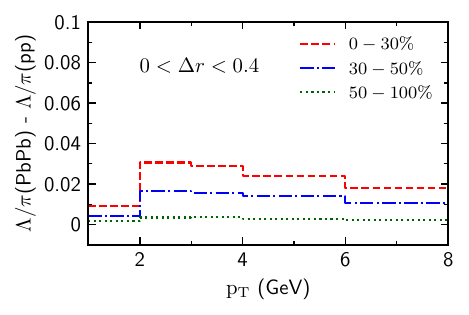}
		\includegraphics[width=0.49\linewidth]{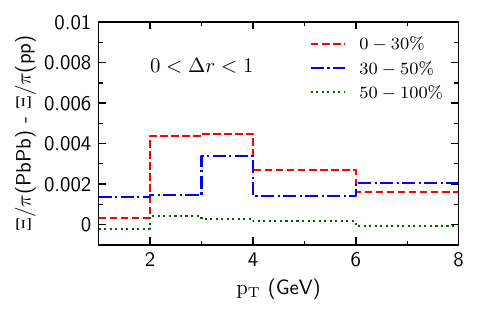}
		\includegraphics[width=0.49\linewidth]{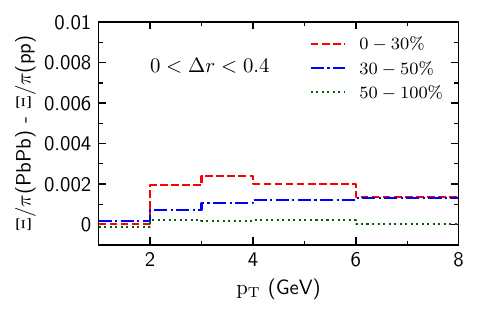}
		\includegraphics[width=0.49\linewidth]{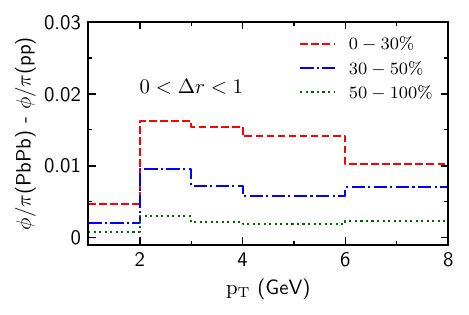}
		\includegraphics[width=0.49\linewidth]{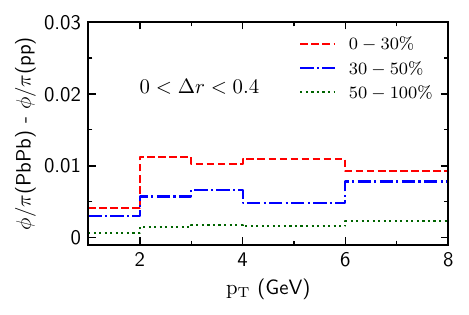}
		\caption{(Color online) The $p_\mathrm{T}$ dependence of the enhancement of  $K/\pi$, $\Lambda/\pi$, $\Xi/\pi$, and $\phi/\pi$ ratios around quenched jets in different centralities of Pb+Pb at $\sqrt{s_\mathrm{NN}} = 5.02$~TeV, relative to their $p+p$ baseline, where the left panels are for identified particles within a distance of ${\Delta r<1}$ from jet axis, and the right panels for ${\Delta r<0.4}$. Jets are selected with ${p^{\rm jet}_\mathrm{T}} > 120$~GeV, $R = 0.4$, and $| \eta_{\rm jet} | < 1.6$.
		}
		\label{fig:dNdpT}
\end{figure*}
	
\begin{figure*}[t!]
		\includegraphics[width=0.49\linewidth]{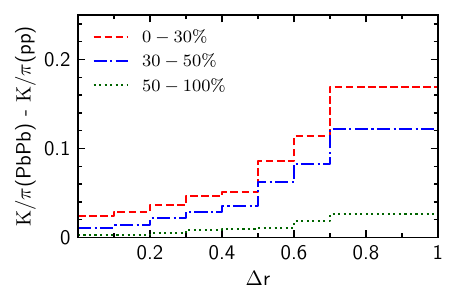}
		\includegraphics[width=0.49\linewidth]{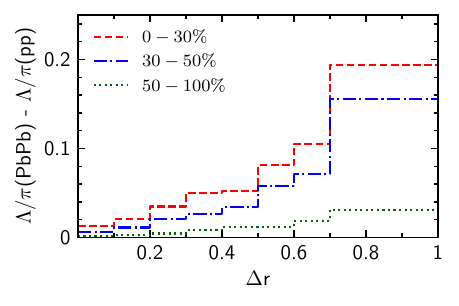}
		\includegraphics[width=0.49\linewidth]{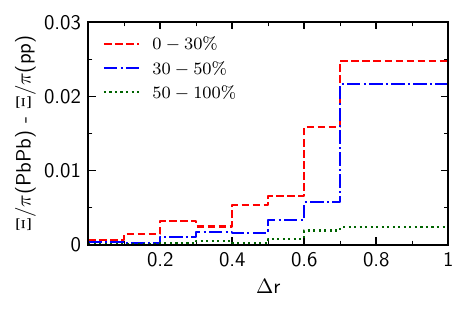}
		\includegraphics[width=0.49\linewidth]{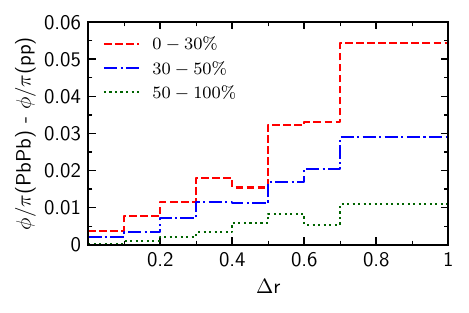}
		\caption{(Color online) The radial distance dependence of the enhancement of $K/\pi$, $\Lambda/\pi$, $\Xi/\pi$, and $\phi/\pi$ ratios around quenched jets in different centralities of Pb+Pb at $\sqrt{s_\mathrm{NN}} = 5.02$~TeV, relative to their $p+p$ baseline. Jets are selected with ${p^{\rm jet}_\mathrm{T}} > 120$~GeV, $R = 0.4$, and $| \eta_{\rm jet} | < 1.6$, and identified particles are selected with $p_\mathrm{T} =2$-4~GeV.}
		\label{fig:dNdr}
\end{figure*}
	
\begin{figure*}
		\includegraphics[width=0.49\linewidth]{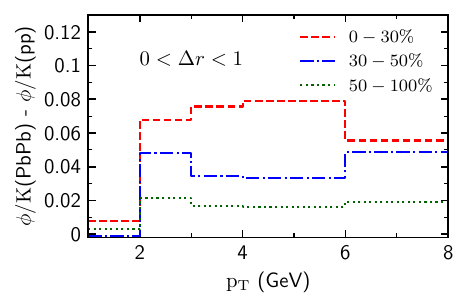}
		\includegraphics[width=0.49\linewidth]{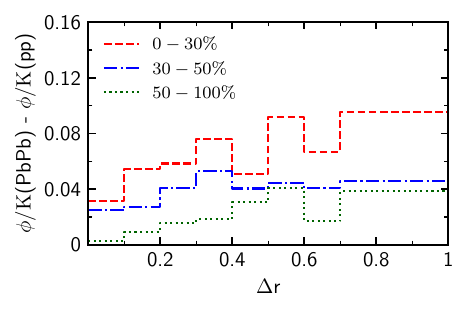}
		\caption{(Color online) The difference of $\phi/K$ ratio around the quenched jets between Pb+Pb and $p+p$ collisions at $\sqrt{s_\mathrm{NN}} = 5.02$~TeV, where the left panel is for the dependence on the $p_\mathrm{T}$ of identified particles within the radial distance of $\Delta r<1$ from the jet axis, and right panel for the dependence on the $\Delta r$ of identified particles within the $p_\mathrm{T}$ region of 2-4~GeV. Jets are selected with ${p^{\rm jet}_\mathrm{T}} > 120$~GeV, $R = 0.4$, and $| \eta_{\rm jet} | < 1.6$.}
		\label{fig:phi_to_k}
\end{figure*}

To illustrate the correction from this procedure, we present in Fig.~\ref{fig:mixed_event} the two-dimensional distributions of kaons (as an example) with respect to jets in 0-30\% Pb+Pb collisions at $\sqrt{s_\mathrm{NN}} = 5.02$~TeV. Kaons are selected within the range of $2 < p_\mathrm{T} <3$ GeV. In the upper left panel, we first calculate the signal jet-kaon pair distribution $S(\Delta \eta, \Delta \phi)$. Around $\Delta \eta\sim 0$, one can observe a prominent jet-like peak on the near side ($\Delta\phi\sim 0$), together with a much smeared peak on the away side ($\Delta\phi\sim \pi$) from the sub-leading jet in a dijet event. These structures are on top of a triangle-like background in the $\Delta\eta$-$\Delta\phi$ plane due to the limited acceptance effect. In the upper right panel, we construct the mixed-event distribution $ME(\Delta \eta, \Delta \phi)$ by pairing  triggered jets and kaons taken from different events, normalized by its value at $\Delta \eta = \Delta \phi =0$. One can see a non-uniform distribution of this mixed-event distribution, resulting from the limited acceptance effect. Shown in the lower left panel is the ratio between the distribution in the upper left panel and that in the upper right panel, or the acceptance-corrected distribution of kaons per trigger jet as defined in Eq.~(\ref{eq:dN_per_jet}).

The distribution in the lower left panel of Fig.~\ref{fig:mixed_event} still contains uncorrelated background and long-range correlations originating from the QGP expansion. We use the side-band method to further remove this background. Following the approach implemented in the CMS experiment~\cite{CMS:2016qnj}, we estimate the background contribution $B(\Delta \eta, \Delta \phi)$ using the $\Delta \phi$ distribution averaged over the range of $1.5 < |\Delta \eta| < 2.5$. After subtracting this background, we obtain our final jet-kaon correlation in the lower right panel of Fig.~\ref{fig:mixed_event}.

The above procedure obtains the two-dimensional distribution $dN/(d\Delta\phi d\Delta \eta)$ within a given $p_\mathrm{T}$ interval.
To achieve the $p_\mathrm{T}$ dependence of the jet-correlated particle yields, we repeat the above analysis for different species of hadrons within various $p_\mathrm{T}$ bins defined with the boundaries of 1, 2, 3, 4, 6, and 8~GeV.
By dividing the distribution (with uncorrelated background subtracted) by the width of each bin, we get the three-dimensional distribution ${d^3N}/{(dp_\mathrm{T} d\Delta \eta d\Delta \phi)}$ of identified particles correlated to quenched jets.

Using this three-dimensional distribution, the $p_\mathrm{T}$ dependence of the jet-correlated particle yields within a given angular region ($\Delta \tilde{r}$) is then given by
	\begin{align}
		\frac{dN}{dp_\mathrm{T}}  \bigg| _{\Delta r < \Delta \tilde{r}} &= \int d\Delta \phi \int d\Delta \eta
		\frac{d^3 N}{dp_\mathrm{T} d\Delta \phi d\Delta \eta } \nonumber\\
		& \times \theta\left(\Delta \tilde{r}-\sqrt{(\Delta\phi)^2 + (\Delta\eta)^2}\right),
	\end{align}
in which the radial distance between an identified particle and the jet axis is defined with $\Delta r=\sqrt{(\Delta\phi)^2 + (\Delta\eta)^2}$ and $\theta$ denotes a step function.

In addition, we will examine the radial distribution of identified particles relative to the jet axis, which can be written as
	\begin{align}
		\frac{dN}{d\Delta r} &= \int dp_\mathrm{T} \int d\Delta \phi \int d\Delta \eta
		\frac{d^3 N}{dp_\mathrm{T} d\Delta \phi d\Delta \eta }
		\nonumber\\
		& \times \delta\left(\Delta r - \sqrt{(\Delta\phi)^2 + (\Delta\eta)^2}\right).
	\end{align}
In practice, this radial distribution will be calculated in annular rings with width of $0.1$ at discrete values of $\Delta r$.

\section{Strangeness enhancement around quenched jets}
\label{sec:strangeness}

We start with the $p_\mathrm{T}$ dependence of strangeness production around jets within a given radial distance from the jet axis. The strangeness production is characterized by ratios of various types of strange particles ($K$, $\Lambda$, $\Xi$, and  $\phi$) to pions. Shown in Fig.~\ref{fig:dNdpT} is the difference of these ratios between Pb+Pb and $p+p$ collisions at $\sqrt{s_\mathrm{NN}} = 5.02$~TeV, left column for the radial distance within $\Delta r < 1$ (around jets) and the right column for $\Delta r < 0.4$ (inside jets). In each panel, results in different centrality regions of Pb+Pb collisions (0-30\%, 30-50\%,  50-100\%) are analyzed. To increase the statistical significance, we average results over hadrons with the same species but different charges, e.g., $\pi$ denotes $(\pi^+ + \pi^- + \pi^0)/3$,  $K$ denotes $(K^+ + K^- + K_S^0)/3$, $\Lambda$ denotes $(\Lambda + \bar{\Lambda})/2$, and $\Xi$ denotes $(\Xi + \bar{\Xi})/2$.

In each panel of Fig.~\ref{fig:dNdpT}, one can clearly observe stronger enhancement of strange-to-non-strange-hadron ratios inside (around) quenched jets in more central Pb+Pb collisions. This can be understood with the stronger jet-medium interactions in more central collisions, which correlate more strangeness rich medium particles to jets through medium response.

Comparing between the left and right columns, we see the strangeness enhancement with ${\Delta r<1}$ is more prominent than that with ${\Delta r<0.4}$, indicating the medium response signal is stronger outside the jet cone than inside the jet cone. This is in line with the expectation that jet-medium interactions transport jet energy from small to large angles relative to the jet axis, and agrees with the findings from other jet observables like the jet shape~\cite{Tachibana:2017syd,Luo:2018pto,Luo:2021hoo,CMS:2018zze}, the inclusive particle correlation with jets~\cite{CMS:2016qnj,ATLAS:2019pid}, and the baryon-to-meson enhancement around quenched jets~\cite{Luo:2021voy}.

In central collisions where jet-induced medium excitation is strong, most strange-to-non-strange-hadron ratios ($\Lambda/\pi$, $\Xi/\pi$, and $\phi/\pi$) get maximum enhancement in the intermediate $p_\mathrm{T}$ (2-4~GeV) region in Pb+Pb collisions relative to their $p+p$ baselines. This can be understood with the parton coalescence process that combines lower $p_\mathrm{T}$ partons into higher $p_\mathrm{T}$ hadrons, and heavier hadrons can inherit stronger radial flow of the QGP and therefore possess higher $p_\mathrm{T}$. Since $\Lambda$, $\Xi$, and $\phi$ have much larger masses than $\pi$, their ratios peak at intermediate $p_\mathrm{T}$. On the other hand, such peak is not seen in the $K/\pi$ ratio since the mass of $K$ is not significantly larger than the mass of $\pi$. Considering that the current AMPT model only includes the coalescence mechanism in hadron formation, its prediction on high-$p_\mathrm{T}$  hadrons ($>6\sim 8$~GeV, depending on the hadron mass) is not quantitatively reliable, since their formation should be dominated by the fragmentation mechanism. This is left for our future improvement.

In Fig.~\ref{fig:dNdr}, we focus on the intermediate $p_\mathrm{T}$ (2-4~GeV) regime of identified particles, where the strongest enhancement is observed for most strange hadrons, and investigate the radial distance ($\Delta r$) dependence of the $K/\pi$, $\Lambda/\pi$, $\Xi/\pi$, and $\phi/\pi$ ratios around quenched jets in Pb+Pb vs. $p+p$ collisions at $\sqrt{s_\mathrm{NN}} = 5.02$~TeV.
Here, one can clearly observe that as the annular bin gets further away from the jet axis, the strangeness enhancement gets stronger. This is caused by the diffusion of jet energy from small to large angles with respect to jet axes, consistent with the conclusion drawn from comparing between the left and right columns of Fig.~\ref{fig:dNdpT}. In addition, due to stronger jet-medium interactions in more central Pb+Pb collisions, an increasing enhancement of the strange-to-non-strange-hadron ratios can be clearly seen from peripheral (50-100\%) to central (0-30\%) collisions.

Apart from strangeness enhancement from jet-medium interaction, the ratios of $\Lambda/\pi$ and $\Xi/\pi$ presented in Figs.~\ref{fig:dNdpT} and~\ref{fig:dNdr} can also be affected by the enhancement of baryon-to-meson production around the quenched jet via the coalescence process. As a complementary demonstration of strangeness enhancement, we study the double-to-single-strange-hadron ratio ($\phi/K$) in Fig.~\ref{fig:phi_to_k}. In the left panel, within the $0<\Delta r<1$ region for the radial distance, we see an enhancement of the $\phi/K$ ratio in Pb+Pb collisions relative to its $p+p$ baseline, which appears more prominent in more central collisions and is strongest in the $p_\mathrm{T}$ region of 2-4~GeV. These features are the same as those observed for the strange-baryon-to-pion ratios previously shown in Fig.~\ref{fig:dNdpT}. In the right panel of Fig.~\ref{fig:phi_to_k}, within a given $p_\mathrm{T}$ regime of the identified particles, we see the difference of the $\phi/K$ ratio between Pb+Pb and $p+p$ collisions increases with $\Delta r$ and appears larger in more central collisions, similar to the ratios between strange and non-strange hadrons earlier.

\begin{figure}
	\includegraphics[width=1\linewidth]{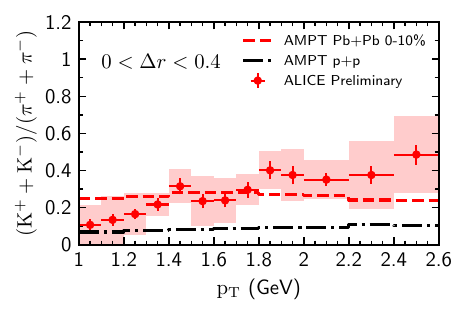}
	\caption{(Color online) The $p_\mathrm{T}$ dependence of the $({K^+} + {K^{-}})/({\pi^+} + \mathrm{\pi^-})$ ratio within a distance of ${\Delta r<0.4}$ from the jet axis in $p+p$ and 0-10\% Pb+Pb collisions at $\sqrt{s_\mathrm{NN}} = 5.02$~TeV, compared to the ALICE preliminary data for Pb+Pb collisions~\cite{ALI-PREL-582548}. Jets are selected with $ 60<{p^{\rm jet}_\mathrm{T}}<140$~GeV, $R = 0.4$, and $| \eta_{\rm jet} | < 0.5$.}
	\label{fig:k_to_pi_ALICE}
\end{figure}


Recently, the ALICE Collaboration has measured the $({K^+} + {K^{-}})/({\pi^+} + \mathrm{\pi^-})$ ratio within quenched jets in 0-10\% Pb+Pb collisions at $\sqrt{s_\mathrm{NN}}=5.02$~TeV~\cite{ALI-PREL-582548}. In Fig.~\ref{fig:k_to_pi_ALICE}, we present our AMPT results using the same experimental setups for both $p+p$ and Pb+Pb collisions. Our results for 0-10\% Pb+Pb collisions reasonably agree with the ALICE data within the experimental uncertainties. We look forward to similar measurements in $p+p$ collisions, in order to verify this strangeness enhancement phenomenon around quenched jets due to medium response.

\section{Summary}
\label{sec:summary}
	
We study the enhancement of strangeness production around quenched jets as a novel signature of jet-induced medium excitation.
When jets traverse the thermalized QGP, the medium constituents can be excited by the hard jets.
These excited partons further thermalize via interacting with medium partons, and thus tend to be richer in strangeness than jet shower partons.
This leads to an access of the strange hadron production correlated with quenched jets through the QGP, compared to jets produced in $p+p$ collisions.

Based on the AMPT simulation, we utilize the jet-particle correlation method to investigate how jet-induced medium excitation affects the final-state chemical compositions of particles produced around quenched jets in heavy-ion collisions. Specifically, we compute the ($\Delta \eta, \Delta \phi$) distributions of various strange hadrons ($K$, $\Lambda$, $\Xi$, and $\phi$) and non-strange pions ($\pi$) with respect to the triggered jet direction in $p+p$ and Pb+Pb collisions at $\sqrt{s_\mathrm{NN}} = 5.02$~TeV. To obtain the final distributions of identified particles correlated with jets, the mixed-event method is used for eliminating the finite acceptance effect and the side-band method is employed for subtracting the uncorrelated background and the long-range correlations. Our results show that the strange hadrons correlated with quenched jets in heavy-ion collisions, quantified by the $K/\pi$, $\Lambda/\pi$, $\Xi/\pi$, and $\phi/\pi$ ratios, are larger than those correlated with vacuum jets in $p+p$ collisions. This enhancement is more pronounced in more central Pb+Pb collisions due to stronger jet-medium interactions there, and is most significant at intermediate $p_\mathrm{T}$, except for the $K/\pi$ ratio, due to the parton coalescence mechanism in hadron formation. Additionally, together with the diffusion of jet energy from small to large angles, the strangeness enhancement is found to increase with the radial distance relative to the jet axis. These features are consistent with signals of medium response found in other jet observables, and are further confirmed with the double-to-single-strange-hadron ($\phi/K$) ratio in our calculation. Therefore, the enhancement of strange hadrons correlated with jets, if confirmed,  could serve as an additional unambiguous signature of medium response, and help further refine our knowledge on the transport properties of the QGP and the hadronization mechanism on the boundary of a color deconfined QGP.
	
It is worth noting that without medium response and parton coalescence, medium modification on the color flow of parton showers can also change the chemical components of hadrons associated with jets~\cite{Sapeta:2007ad}. However, its dependence on the $p_\mathrm{T}$ and $\Delta r$ of hadrons would appear different from our predictions in this work. Furthermore, the current AMPT model does not account for the medium-induced gluon emission process, which could influence the magnitude of both jet quenching and jet-induced medium excitation~\cite{Qin:2009uh, Neufeld:2009ep}. However, since the strangeness enhancement inside a hot QCD medium and the modification of the chemical composition of hadrons via parton coalescence are general features of relativistic heavy-ion collisions that are independent of detailed model implementations, our qualitative prediction on the strangeness enhancement around quenched jets should be robust and can be readily verified by experimental measurements in the near future.

\section*{Acknowledgements}
We are grateful to Yaxian Mao for helpful discussions. This work is supported in part by the China Postdoctoral Science Foundation under Grant No.~2024M751790 (AL), and in part by the National Natural Science Foundation of China (NSFC) under Grant Nos.~12347150 (AL), 12175122 (SC, AL), 2021-867 (SC, AL), 12225503 (GYQ) and 11935007 (GYQ). Some of the calculations were performed in the Nuclear Science Computing Center at Central China Normal University (NSC3), Wuhan, Hubei, China.

\bibliographystyle{h-physrev5} 
\bibliography{references}   

\begin{thebibliography}{100}

\bibitem{Gyulassy:2004zy}
M.~Gyulassy and L.~McLerran,
\newblock Nucl. Phys. A {\bf 750}, 30 (2005), arXiv:nucl-th/0405013.

\bibitem{Jacobs:2004qv}
P.~Jacobs and X.-N. Wang,
\newblock Prog. Part. Nucl. Phys. {\bf 54}, 443 (2005), arXiv:hep-ph/0405125.

\bibitem{Busza:2018rrf}
W.~Busza, K.~Rajagopal, and W.~van~der Schee,
\newblock Ann. Rev. Nucl. Part. Sci. {\bf 68}, 339 (2018), arXiv:1802.04801.

\bibitem{Elfner:2022iae}
H.~Elfner and B.~M\"uller,
\newblock J. Phys. G {\bf 50}, 103001 (2023), arXiv:2210.12056.

\bibitem{Wang:1992qdg}
X.-N. Wang and M.~Gyulassy,
\newblock Phys. Rev. Lett. {\bf 68}, 1480 (1992).

\bibitem{Bass:2008rv}
S.~A. Bass {\em et~al.},
\newblock Phys. Rev. C {\bf 79}, 024901 (2009), arXiv:0808.0908.

\bibitem{Qin:2015srf}
G.-Y. Qin and X.-N. Wang,
\newblock Int. J. Mod. Phys. E {\bf 24}, 1530014 (2015), arXiv:1511.00790.

\bibitem{Majumder:2010qh}
A.~Majumder and M.~Van~Leeuwen,
\newblock Prog. Part. Nucl. Phys. {\bf 66}, 41 (2011), arXiv:1002.2206.

\bibitem{Blaizot:2015lma}
J.-P. Blaizot and Y.~Mehtar-Tani,
\newblock Int. J. Mod. Phys. E {\bf 24}, 1530012 (2015), arXiv:1503.05958.

\bibitem{Cao:2020wlm}
S.~Cao and X.-N. Wang,
\newblock Rept. Prog. Phys. {\bf 84}, 024301 (2021), arXiv:2002.04028.

\bibitem{Qin:2010mn}
G.-Y. Qin and B.~Muller,
\newblock Phys. Rev. Lett. {\bf 106}, 162302 (2011), arXiv:1012.5280,
\newblock [Erratum: Phys.Rev.Lett. 108, 189904 (2012)].

\bibitem{Majumder:2011uk}
A.~Majumder and C.~Shen,
\newblock Phys. Rev. Lett. {\bf 109}, 202301 (2012), arXiv:1103.0809.

\bibitem{Dai:2012am}
W.~Dai, I.~Vitev, and B.-W. Zhang,
\newblock Phys. Rev. Lett. {\bf 110}, 142001 (2013), arXiv:1207.5177.

\bibitem{Blaizot:2013hx}
J.-P. Blaizot, E.~Iancu, and Y.~Mehtar-Tani,
\newblock Phys. Rev. Lett. {\bf 111}, 052001 (2013), arXiv:1301.6102.

\bibitem{Chien:2016led}
Y.-T. Chien and I.~Vitev,
\newblock Phys. Rev. Lett. {\bf 119}, 112301 (2017), arXiv:1608.07283.

\bibitem{Caucal:2018dla}
P.~Caucal, E.~Iancu, A.~H. Mueller, and G.~Soyez,
\newblock Phys. Rev. Lett. {\bf 120}, 232001 (2018), arXiv:1801.09703.

\bibitem{Xing:2019xae}
W.-J. Xing, S.~Cao, G.-Y. Qin, and H.~Xing,
\newblock Phys. Lett. B {\bf 805}, 135424 (2020), arXiv:1906.00413.

\bibitem{Huss:2020dwe}
A.~Huss {\em et~al.},
\newblock Phys. Rev. Lett. {\bf 126}, 192301 (2021), arXiv:2007.13754.

\bibitem{Mehtar-Tani:2021fud}
Y.~Mehtar-Tani, D.~Pablos, and K.~Tywoniuk,
\newblock Phys. Rev. Lett. {\bf 127}, 252301 (2021), arXiv:2101.01742.

\bibitem{Zhao:2021vmu}
W.~Zhao, W.~Ke, W.~Chen, T.~Luo, and X.-N. Wang,
\newblock Phys. Rev. Lett. {\bf 128}, 022302 (2022), arXiv:2103.14657.

\bibitem{Cao:2024pxc}
S.~Cao, A.~Majumder, R.~Modarresi-Yazdi, I.~Soudi, and Y.~Tachibana,
\newblock Int. J. Mod. Phys. E {\bf 33}, 2430002 (2024), arXiv:2401.10026.

\bibitem{Schenke:2009gb}
B.~Schenke, C.~Gale, and S.~Jeon,
\newblock Phys. Rev. C {\bf 80}, 054913 (2009), arXiv:0909.2037.

\bibitem{Zapp:2013vla}
K.~C. Zapp,
\newblock Eur. Phys. J. C {\bf 74}, 2762 (2014), arXiv:1311.0048.

\bibitem{Casalderrey-Solana:2014bpa}
J.~Casalderrey-Solana, D.~C. Gulhan, J.~G. Milhano, D.~Pablos, and
  K.~Rajagopal,
\newblock JHEP {\bf 10}, 019 (2014), arXiv:1405.3864,
\newblock [Erratum: JHEP 09, 175 (2015)].

\bibitem{Cao:2017qpx}
S.~Cao and A.~Majumder,
\newblock Phys. Rev. C {\bf 101}, 024903 (2020), arXiv:1712.10055.

\bibitem{JETSCAPE:2017eso}
JETSCAPE, S.~Cao {\em et~al.},
\newblock Phys. Rev. C {\bf 96}, 024909 (2017), arXiv:1705.00050.

\bibitem{Putschke:2019yrg}
J.~H. Putschke {\em et~al.},
\newblock (2019), arXiv:1903.07706.

\bibitem{Luo:2023nsi}
T.~Luo, Y.~He, S.~Cao, and X.-N. Wang,
\newblock Phys. Rev. C {\bf 109}, 034919 (2024), arXiv:2306.13742.

\bibitem{Karpenko:2024fgg}
I.~Karpenko, A.~Lind, M.~Rohrmoser, J.~Aichelin, and P.-B. Gossiaux,
\newblock (2024), arXiv:2404.14579.

\bibitem{JET:2013cls}
JET, K.~M. Burke {\em et~al.},
\newblock Phys. Rev. C {\bf 90}, 014909 (2014), arXiv:1312.5003.

\bibitem{JETSCAPE:2021ehl}
JETSCAPE, S.~Cao {\em et~al.},
\newblock Phys. Rev. C {\bf 104}, 024905 (2021), arXiv:2102.11337.

\bibitem{Xie:2022ght}
M.~Xie, W.~Ke, H.~Zhang, and X.-N. Wang,
\newblock Phys. Rev. C {\bf 108}, L011901 (2023), arXiv:2206.01340.

\bibitem{Chen:2024epd}
B.~Chen, X.~Chen, X.~Li, Z.-R. Zhu, and K.~Zhou,
\newblock (2024), arXiv:2404.18217.

\bibitem{Karmakar:2024jak}
B.~Karmakar {\em et~al.},
\newblock Phys. Rev. C {\bf 110}, 044906 (2024), arXiv:2403.17817.

\bibitem{Karmakar:2023ity}
B.~Karmakar {\em et~al.},
\newblock Phys. Rev. C {\bf 108}, 044907 (2023), arXiv:2305.11318.

\bibitem{Liu:2023rfi}
F.-L. Liu, X.-Y. Wu, S.~Cao, G.-Y. Qin, and X.-N. Wang,
\newblock Phys. Lett. B {\bf 848}, 138355 (2024), arXiv:2304.08787.

\bibitem{Cao:2022odi}
S.~Cao and G.-Y. Qin,
\newblock Ann. Rev. Nucl. Part. Sci. {\bf 73}, 205 (2023), arXiv:2211.16821.

\bibitem{He:2018xjv}
Y.~He {\em et~al.},
\newblock Phys. Rev. C {\bf 99}, 054911 (2019), arXiv:1809.02525.

\bibitem{JETSCAPE:2022jer}
JETSCAPE, A.~Kumar {\em et~al.},
\newblock Phys. Rev. C {\bf 107}, 034911 (2023), arXiv:2204.01163.

\bibitem{He:2022evt}
Y.~He {\em et~al.},
\newblock Phys. Rev. C {\bf 106}, 044904 (2022), arXiv:2201.08408.

\bibitem{Casalderrey-Solana:2016jvj}
J.~Casalderrey-Solana, D.~Gulhan, G.~Milhano, D.~Pablos, and K.~Rajagopal,
\newblock JHEP {\bf 03}, 135 (2017), arXiv:1609.05842.

\bibitem{Tachibana:2017syd}
Y.~Tachibana, N.-B. Chang, and G.-Y. Qin,
\newblock Phys. Rev. C {\bf 95}, 044909 (2017), arXiv:1701.07951.

\bibitem{KunnawalkamElayavalli:2017hxo}
R.~Kunnawalkam~Elayavalli and K.~C. Zapp,
\newblock JHEP {\bf 07}, 141 (2017), arXiv:1707.01539.

\bibitem{Milhano:2017nzm}
G.~Milhano, U.~A. Wiedemann, and K.~C. Zapp,
\newblock Phys. Lett. B {\bf 779}, 409 (2018), arXiv:1707.04142.

\bibitem{Chen:2020tbl}
W.~Chen, S.~Cao, T.~Luo, L.-G. Pang, and X.-N. Wang,
\newblock Phys. Lett. B {\bf 810}, 135783 (2020), arXiv:2005.09678.

\bibitem{Park:2018acg}
C.~Park, S.~Jeon, and C.~Gale,
\newblock Nucl. Phys. A {\bf 982}, 643 (2019), arXiv:1807.06550.

\bibitem{Luo:2018pto}
T.~Luo, S.~Cao, Y.~He, and X.-N. Wang,
\newblock Phys. Lett. B {\bf 782}, 707 (2018), arXiv:1803.06785.

\bibitem{Casalderrey-Solana:2019ubu}
J.~Casalderrey-Solana, G.~Milhano, D.~Pablos, and K.~Rajagopal,
\newblock JHEP {\bf 01}, 044 (2020), arXiv:1907.11248.

\bibitem{Chang:2019sae}
N.-B. Chang, Y.~Tachibana, and G.-Y. Qin,
\newblock Phys. Lett. B {\bf 801}, 135181 (2020), arXiv:1906.09562.

\bibitem{Tachibana:2020mtb}
Y.~Tachibana, C.~Shen, and A.~Majumder,
\newblock Phys. Rev. C {\bf 106}, L021902 (2022), arXiv:2001.08321.

\bibitem{Yang:2023dwc}
Z.~Yang, Y.~He, I.~Moult, and X.-N. Wang,
\newblock Phys. Rev. Lett. {\bf 132}, 011901 (2024), arXiv:2310.01500.

\bibitem{Xing:2024yrb}
W.-J. Xing, S.~Cao, G.-Y. Qin, and X.-N. Wang,
\newblock (2024), arXiv:2409.12843.

\bibitem{Casalderrey-Solana:2004fdk}
J.~Casalderrey-Solana, E.~V. Shuryak, and D.~Teaney,
\newblock J. Phys. Conf. Ser. {\bf 27}, 22 (2005), arXiv:hep-ph/0411315.

\bibitem{Chaudhuri:2005vc}
A.~K. Chaudhuri and U.~Heinz,
\newblock Phys. Rev. Lett. {\bf 97}, 062301 (2006), arXiv:nucl-th/0503028.

\bibitem{Ruppert:2005uz}
J.~Ruppert and B.~Muller,
\newblock Phys. Lett. B {\bf 618}, 123 (2005), arXiv:hep-ph/0503158.

\bibitem{Gubser:2007ga}
S.~S. Gubser, S.~S. Pufu, and A.~Yarom,
\newblock Phys. Rev. Lett. {\bf 100}, 012301 (2008), arXiv:0706.4307.

\bibitem{Chesler:2007an}
P.~M. Chesler and L.~G. Yaffe,
\newblock Phys. Rev. Lett. {\bf 99}, 152001 (2007), arXiv:0706.0368.

\bibitem{Qin:2009uh}
G.~Y. Qin, A.~Majumder, H.~Song, and U.~Heinz,
\newblock Phys. Rev. Lett. {\bf 103}, 152303 (2009), arXiv:0903.2255.

\bibitem{Neufeld:2009ep}
R.~B. Neufeld and B.~Muller,
\newblock Phys. Rev. Lett. {\bf 103}, 042301 (2009), arXiv:0902.2950.

\bibitem{Li:2010ts}
H.~Li, F.~Liu, G.-l. Ma, X.-N. Wang, and Y.~Zhu,
\newblock Phys. Rev. Lett. {\bf 106}, 012301 (2011), arXiv:1006.2893.

\bibitem{Ma:2010dv}
G.-L. Ma and X.-N. Wang,
\newblock Phys. Rev. Lett. {\bf 106}, 162301 (2011), arXiv:1011.5249.

\bibitem{Casalderrey-Solana:2020rsj}
J.~Casalderrey-Solana, J.~G. Milhano, D.~Pablos, K.~Rajagopal, and X.~Yao,
\newblock JHEP {\bf 05}, 230 (2021), arXiv:2010.01140.

\bibitem{Neufeld:2008dx}
R.~B. Neufeld,
\newblock Phys. Rev. C {\bf 79}, 054909 (2009), arXiv:0807.2996.

\bibitem{Bouras:2014rea}
I.~Bouras, B.~Betz, Z.~Xu, and C.~Greiner,
\newblock Phys. Rev. C {\bf 90}, 024904 (2014), arXiv:1401.3019.

\bibitem{Betz:2010qh}
B.~Betz, J.~Noronha, G.~Torrieri, M.~Gyulassy, and D.~H. Rischke,
\newblock Phys. Rev. Lett. {\bf 105}, 222301 (2010), arXiv:1005.5461.

\bibitem{Tachibana:2015qxa}
Y.~Tachibana and T.~Hirano,
\newblock Phys. Rev. C {\bf 93}, 054907 (2016), arXiv:1510.06966.

\bibitem{Neufeld:2011yh}
R.~B. Neufeld and I.~Vitev,
\newblock Phys. Rev. C {\bf 86}, 024905 (2012), arXiv:1105.2067.

\bibitem{Renk:2013pua}
T.~Renk,
\newblock Phys. Rev. C {\bf 88}, 044905 (2013), arXiv:1306.2739.

\bibitem{Betz:2008ka}
B.~Betz {\em et~al.},
\newblock Phys. Rev. C {\bf 79}, 034902 (2009), arXiv:0812.4401.

\bibitem{Chen:2017zte}
W.~Chen, S.~Cao, T.~Luo, L.-G. Pang, and X.-N. Wang,
\newblock Phys. Lett. B {\bf 777}, 86 (2018), arXiv:1704.03648.

\bibitem{Yang:2021qtl}
Z.~Yang {\em et~al.},
\newblock Phys. Rev. Lett. {\bf 127}, 082301 (2021), arXiv:2101.05422.

\bibitem{Yang:2022nei}
Z.~Yang, T.~Luo, W.~Chen, L.-G. Pang, and X.-N. Wang,
\newblock Phys. Rev. Lett. {\bf 130}, 052301 (2023), arXiv:2203.03683.

\bibitem{CMS:CMS-PAS-HIN-23-006}
CMS,
\newblock CMS-PAS-HIN-23-006  (2024).

\bibitem{Chen:2021rrp}
W.~Chen, S.~Cao, T.~Luo, L.-G. Pang, and X.-N. Wang,
\newblock Nucl. Phys. A {\bf 1005}, 121934 (2021).

\bibitem{Luo:2021voy}
A.~Luo, Y.-X. Mao, G.-Y. Qin, E.-K. Wang, and H.-Z. Zhang,
\newblock Phys. Lett. B {\bf 837}, 137638 (2023), arXiv:2109.14314.

\bibitem{Sirimanna:2022zje}
C.~Sirimanna {\em et~al.},
\newblock Phys. Rev. C {\bf 108}, 014911 (2023), arXiv:2211.15553.

\bibitem{Dale-Gau:2023ree}
STAR, G.~Dale-Gau,
\newblock PoS {\bf HardProbes2023}, 173 (2024), arXiv:2312.11362.

\bibitem{Rafelski:1982pu}
J.~Rafelski and B.~Muller,
\newblock Phys. Rev. Lett. {\bf 48}, 1066 (1982),
\newblock [Erratum: Phys.Rev.Lett. 56, 2334 (1986)].

\bibitem{Fries:2003kq}
R.~J. Fries, B.~Muller, C.~Nonaka, and S.~A. Bass,
\newblock Phys. Rev. C {\bf 68}, 044902 (2003), arXiv:nucl-th/0306027.

\bibitem{Fries:2003vb}
R.~J. Fries, B.~Muller, C.~Nonaka, and S.~A. Bass,
\newblock Phys. Rev. Lett. {\bf 90}, 202303 (2003), arXiv:nucl-th/0301087.

\bibitem{Greco:2003mm}
V.~Greco, C.~M. Ko, and P.~Levai,
\newblock Phys. Rev. C {\bf 68}, 034904 (2003), arXiv:nucl-th/0305024.

\bibitem{Lin:2004en}
Z.-W. Lin, C.~M. Ko, B.-A. Li, B.~Zhang, and S.~Pal,
\newblock Phys. Rev. C {\bf 72}, 064901 (2005), arXiv:nucl-th/0411110.

\bibitem{Zhang:2005ni}
B.~Zhang, L.-W. Chen, and C.-M. Ko,
\newblock Phys. Rev. C {\bf 72}, 024906 (2005), arXiv:nucl-th/0502056.

\bibitem{Lin:2001zk}
Z.-W. Lin and C.~M. Ko,
\newblock Phys. Rev. C {\bf 65}, 034904 (2002), arXiv:nucl-th/0108039.

\bibitem{Chen:2004dv}
L.-W. Chen, C.~M. Ko, and Z.-W. Lin,
\newblock Phys. Rev. C {\bf 69}, 031901 (2004), arXiv:nucl-th/0312124.

\bibitem{Xu:2011fe}
J.~Xu and C.~M. Ko,
\newblock Phys. Rev. C {\bf 84}, 014903 (2011), arXiv:1103.5187.

\bibitem{Ma:2013pha}
G.-L. Ma,
\newblock Phys. Rev. C {\bf 87}, 064901 (2013), arXiv:1304.2841.

\bibitem{Ma:2013bia}
G.-L. Ma,
\newblock Phys. Lett. B {\bf 724}, 278 (2013), arXiv:1302.5873.

\bibitem{Ma:2013gga}
G.-L. Ma,
\newblock Phys. Rev. C {\bf 88}, 021902 (2013), arXiv:1306.1306.

\bibitem{Ma:2013uqa}
G.-L. Ma,
\newblock Phys. Rev. C {\bf 89}, 024902 (2014), arXiv:1309.5555.

\bibitem{Wang:1991hta}
X.-N. Wang and M.~Gyulassy,
\newblock Phys. Rev. D {\bf 44}, 3501 (1991).

\bibitem{Gyulassy:1994ew}
M.~Gyulassy and X.-N. Wang,
\newblock Comput. Phys. Commun. {\bf 83}, 307 (1994), arXiv:nucl-th/9502021.

\bibitem{Zhang:1997ej}
B.~Zhang,
\newblock Comput. Phys. Commun. {\bf 109}, 193 (1998), arXiv:nucl-th/9709009.

\bibitem{Gao:2016ldo}
Z.~Gao, A.~Luo, G.-L. Ma, G.-Y. Qin, and H.-Z. Zhang,
\newblock Phys. Rev. C {\bf 97}, 044903 (2018), arXiv:1612.02548.

\bibitem{Luo:2021hoo}
A.~Luo, Y.-X. Mao, G.-Y. Qin, E.-K. Wang, and H.-Z. Zhang,
\newblock Eur. Phys. J. C {\bf 82}, 156 (2022), arXiv:2107.11751.

\bibitem{Li:1995pra}
B.-A. Li and C.~M. Ko,
\newblock Phys. Rev. C {\bf 52}, 2037 (1995), arXiv:nucl-th/9505016.

\bibitem{CMS:2016cvr}
CMS, V.~Khachatryan {\em et~al.},
\newblock JHEP {\bf 11}, 055 (2016), arXiv:1609.02466.

\bibitem{CMS:2016qnj}
CMS, V.~Khachatryan {\em et~al.},
\newblock JHEP {\bf 02}, 156 (2016), arXiv:1601.00079.

\bibitem{CMS:2018zze}
CMS, A.~M. Sirunyan {\em et~al.},
\newblock JHEP {\bf 05}, 006 (2018), arXiv:1803.00042.

\bibitem{CMS:2021nhn}
CMS, A.~M. Sirunyan {\em et~al.},
\newblock JHEP {\bf 05}, 116 (2021), arXiv:2101.04720.

\bibitem{Cacciari:2008gp}
M.~Cacciari, G.~P. Salam, and G.~Soyez,
\newblock JHEP {\bf 04}, 063 (2008), arXiv:0802.1189.

\bibitem{Cacciari:2011ma}
M.~Cacciari, G.~P. Salam, and G.~Soyez,
\newblock Eur. Phys. J. C {\bf 72}, 1896 (2012), arXiv:1111.6097.

\bibitem{ATLAS:2019pid}
ATLAS, G.~Aad {\em et~al.},
\newblock Phys. Rev. C {\bf 100}, 064901 (2019), arXiv:1908.05264,
\newblock [Erratum: Phys.Rev.C 101, 059903 (2020)].

\bibitem{ALI-PREL-582548}
ALICE,
\newblock ALI-PREL-582548  (2024).

\bibitem{Sapeta:2007ad}
S.~Sapeta and U.~A. Wiedemann,
\newblock Eur. Phys. J. C {\bf 55}, 293 (2008), arXiv:0707.3494.

\end{thebibliography}

\end{document}